\documentclass[twocolumn,aps,showpacs,floatfix]{revtex4}
\usepackage{graphicx}% Include figure files 
\usepackage{bm}% bold math

\begin{document}

\title{1D model for the dynamics and expansion of elongated
Bose-Einstein condensates}

 \author{Pietro Massignan} 
\email{massignan@lens.unifi.it}
  \author{Michele Modugno} 
\email{modugno@fi.infn.it} \affiliation{%
  INFM - LENS - Dipartimento di Fisica, Universit\`a di Firenze\\ Via
  Nello Carrara 1, 50019 Sesto Fiorentino, Italy }%

\date{\today}

\begin{abstract}
We present a 1D effective model for the evolution of a cigar-shaped
Bose-Einstein condensate in time dependent potentials whose radial
component is harmonic.  We apply this model to investigate the
dynamics and expansion of condensates in 1D optical
lattices, by comparing our predictions with recent experimental data
and theoretical results. We also discuss negative-mass effects 
which could be probed during the expansion of a condensate
moving in an optical lattice.
\end{abstract}
 \maketitle

%===============================================================
\section{Introduction}
\label{sec:intro}
%===============================================================

The realization of Bose-Einstein condensates (BECs) in 1D optical
lattices provides an important tool to investigate the superfluid and
coherence properties of macroscopic quantum systems.
Recent experiments have allowed to probe static and dynamic features
of these systems such as the Josephson effect \cite{anderson,cataliotti}, 
the emergence of squeezed states \cite{orzel}, 
the transition from superfluid to
dissipative dynamics \cite{burger}, Bloch oscillations \cite{morsch},
interference patterns produced by freely expanding  condensates
\cite{pedri,morsch2}, and band spectroscopy \cite{rolston}.

These results have stimulated a great interest also from the
theoretical point of view, and several models have been proposed to
interpret the experimental data.  In particular, new analytic results
have been derived in the so called ``tight binding regime'', based on
the assumption that the condensate wave functions in different sites
are well separated \cite{pedri,kraemer,cataliotti}.

A very important point would be to compare these results  with the
numerical solution of the full 3D Gross-Pitaevskii equation (GPE) that
describes the dynamics of BECs at very low temperatures, allowing also for 
the exploration of the weak binding regime.  However, in
general the solution of the 3D GPE is a formidable task that
suffers severe limitations due to the large memory storage and
computational times required to simulate the experimental
configurations.  To overcome these difficulties the GPE is
usually reduced to a 1D equation by means of a static renormalization
of the inter-atomic scattering length \cite{tripp,jackson,chiofalo},
by using the fact that these systems are effectively one-dimensional.
This allows for a realistic description of
the trapped axial dynamics, but at the cost of loosing any information 
on the radial
degrees of freedom. In general this reduction is indeed too rude,
since the transverse dimensions actually play a crucial role both during
the expansion, when the 1D constraint is removed, and during the
trapped dynamics, where the interplay between the axial and radial
components is necessary to account for collective excitations, {\em e.g.}
quadrupole oscillations.

Recently the authors of \cite{npse} have proposed an effective 1D
wave-equation for BECs confined in cylindrically symmetric
potentials, by using a different approach that keeps track also of
the transverse dimensions of the condensate.  The derivation is based
on a variational Ansatz and the wave function of the condensate is
factorized in the product of an axial component and a gaussian radial
one. The axial wave function satisfies a 1D non polynomial
Schr\"odinger equation (NPSE) that gives remarkable results for the
axial ground-state and dynamics of cigar-shaped condensates in
cylindrically symmetric traps.  However, the derivation in \cite{npse}
is based on the assumption of a static harmonic radial confinement,
and therefore the NPSE is not suitable to study the dynamics under
variation of the trapping potential, in particular the ballistic
expansion of the condensate after being released from the trap.

In this paper we present a new model that combines the idea of
gaussian factorization with the unitary scaling and gauge
transformations of \cite{castindum,kagan}, allowing for a suitable
description of both the transverse and 
longitudinal evolution of elongated BECs in the presence of a
time-dependent harmonic potential plus an arbitrary axial component.
The above transformations are used to reabsorb the evolution due to the
time variations of the harmonic potential, the latter being replaced
by an effective confinement that makes feasible a gaussian
factorization even in the case of a sudden release of the external
confinement. The evolution of the system is described by an effective
1D GPE that is \emph{dynamically rescaled}
(\emph{dr}-GPE) by means of the transformations of
\cite{castindum,kagan}. Despite the one dimensional character of this equation,
the model is capable to account for the transverse
dynamics through an algebraic equation for the radial width
coupled to the \emph{dr}-GPE. Therefore this model represents an useful
tool to describe both the dynamics and the expansion of
BECs in elongated geometries.

The Paper is organized as follows. In Section 
\ref{sec:model} we present the derivation of the \emph{dr}-GPE, and in 
Section \ref{sec:harmonic} we discuss its properties in the presence
of a pure harmonic potential, by making a comparison
with analytical results in the Thomas-Fermi limit.
In Section \ref{sec:optical} we use the \emph{dr}-GPE to discuss
the dynamics and expansion of BECs in the presence of 1D optical lattices,
by comparing its predictions with recent experimental and theoretical 
results \cite{cataliotti,pedri,kraemer}. We also present a method to probe
negative-mass effects during the expansion of a condensate in 
the presence of an optical lattice.

%=====================================================
\section{The model}
\label{sec:model}
%=====================================================

We consider a condensate confined in a time-dependent harmonic
trapping potential $U_{ho}$ with an additional axial component
$U_{1D}$
\begin{eqnarray}
U({\bm r},t)&=&U_{ho}({\bm r},t) + U_{1D}(r_z,t)\\
&=&\frac{1}{2}m\omega_\perp^2(t)r_\perp^2+
\frac{1}{2}m\omega_z^2(t)r_z^2 + U_{1D}(r_z,t)\nonumber.
\end{eqnarray}
The Gross-Pitaevskii equation for the wave-function $\Psi$ of the
condensate can be derived from the functional \cite{BEC_review}
\begin{equation}
\label{eq:GP_action}
\!\!\!S\left[\Psi\right]\!=\!\!\int \!\!d t\!\! \int
\!\!d^3\mathbf{r}\ \Psi^*\!
\left[i\hbar\partial_t+\frac{\hbar^2}{2m}\nabla^2-
U-\frac{gN}{2}|\Psi|^2 \right]\!\Psi
\end{equation}
where $N$ is the number of condensed atoms, $g=4\pi\hbar^2a/m$ the
coupling strength, $m$ the atomic mass and $a$ the inter-atomic
scattering length.

In order to account for the time-dependent variations of the confining
potential it is convenient to apply the unitary gauge and scaling
transform of \cite{castindum,kagan}. Therefore we introduce a rescaled
wave function $\tilde{\Psi}({\bf x},t)$, depending on the rescaled
spatial coordinates $x_i\equiv r_i/\lambda_i(t)$ and related to the
true wave function $\Psi({\bf r},t)$ by
\begin{equation}
\label{eq:ansatz_CD}
\Psi(\mathbf{r},t)= e^{\textstyle\frac{im}{2\hbar}\sum r_j^2
 \frac{\dot{\lambda_j}(t)}{\lambda_j(t)}}
 \frac{\tilde{\Psi}\left(\left\{
 {r_j}/{\lambda_j(t)}\right\},t\right)}{
 \sqrt{\lambda_1(t)\lambda_2(t)\lambda_3(t)}}\,.
\end{equation}
The scaling parameters $\lambda_j(t)$ are solutions of
\cite{castindum}
\begin{equation}
\label{eq:lambda}
\ddot{\lambda_j}(t)=
\frac{\omega^2_j(0)}{\lambda_j(t)\lambda_1(t)\lambda_2(t)\lambda_3(t)}
-\omega_j^2(t)\lambda_j(t)
\end{equation}
with the initial conditions ${\bm\lambda}(0)=1$ and
${\dot{\bm\lambda}}(0)=0$.

The use of this scaling equations has two advantages: (i) the
evolution of $\tilde{\Psi}$ due to the variation of the harmonic
trapping potential is mostly absorbed by the scaling and gauge
transform \cite{epjd} (in the Thomas-Fermi limit $|\tilde{\Psi}|$ is
frozen to its initial value); (ii) the rescaled wave function
$\tilde{\Psi}$ always evolves in the presence of a fictitious harmonic
confinement, that depends on the trapping frequencies at $t=0$, and
therefore the gaussian factorization is applicable even in the case of
ballistic expansion, when the real trapping is turned off.

In case of elongated condensates we can obtain a 1D effective model by
factorizing the rescaled wave function $\tilde{\Psi}({\bm x},t)$ in
the product of an axial component $\tilde{\psi}(z,t)$ and a gaussian
radial component $\tilde{\phi}(x,y,t;\sigma(z,t))$, allowing for an
exact integration over the transverse coordinates \cite{jackson,npse}
\begin{equation}
\label{eq:factor}
\tilde{\Psi}({\bf x},t)=\tilde{\phi}\left(x,y,t;\sigma(z,t)\right)
\tilde{\psi}(z,t)
\end{equation} 
\begin{equation}
\tilde{\phi}(x,y;\sigma(z,t))
=\frac{1}{\sqrt{\pi}\sigma}e^{\displaystyle-(x^2+y^2)/2\sigma^2}
\end{equation}
with the normalization conditions
\begin{equation}
\int dz|\tilde{\psi}(z,t)|^2=1= \int\!\!\!\!\int dx
dy\tilde{|\phi}(x,y;\sigma)|^2\,.
\end{equation}
With this choice the transverse size of $\tilde{\Psi}$ is
characterized by the width $\sigma(z,t)$ that is expected to be a
slowly varying function of time, since most of the evolution due to
the variation of the trapping potential is absorbed by the unitary
transform (\ref{eq:ansatz_CD}). We note that from Eq.\
(\ref{eq:ansatz_CD}) and (\ref{eq:factor}) even the true wave function
$\Psi$ can be expressed in the factorized form
\begin{equation}
\Psi\left({\bf
r},t\right)=\phi\left(r_x,r_y,t;\Sigma(r_z,t)\right)\psi(r_z,t)
\end{equation} 
whose transverse width $\Sigma$ is given by
\begin{equation}
\label{eq:gaussian_width}
\Sigma(r_z,t)\equiv\lambda_\perp(t)\sigma(r_z/\lambda_z,t).
\end{equation}

The equations of motion for $\tilde{\psi}$ and $\sigma$ can be derived
by using a variational procedure from the action (\ref{eq:GP_action}),
integrating out the dependence on $\tilde{\phi}$. The problem can be
further simplified in the so called ``slowly varying approximation'',
that is when the contribution given by the axial variations of the radial
wave function $\tilde{\phi}$ is negligible, $\nabla^2 \tilde{\phi}
\approx\nabla^2_\perp\tilde{\phi}$ \cite{npse}.  In this approximation
the action becomes
\begin{widetext}
{\setlength\arraycolsep{0pt}
\begin{eqnarray}
S\left[\Psi\right] \simeq\int\!\!dt\int\!\!dz\ \tilde{\psi}^*(z)
&&\left\{
i\hbar\partial_t+\frac{\hbar^2}{2m}\frac{1}{\lambda_z^2}\nabla^2_z
-\frac{m\omega_z^2(0)}{2}\frac{z^2}{\lambda_z\lambda_\perp^2}
-U_{1D}(\lambda_z z,t) \right.\nonumber\\ &&\left.\ \ \ \ \ \ \ \ \ \
-\frac{\hbar^2}{2m}\frac{1}{\sigma^2\lambda_\perp^2}
-\frac{m\omega_\perp^2(0)}{2}\frac{\sigma^2}{\lambda_z\lambda_\perp^2}
-\frac{1}{2}\frac{gN}{2\pi\sigma^2}\frac{|\tilde{\psi}|^2}
{\lambda_z\lambda_\perp^2} \right\}\tilde{\psi}(z)
\end{eqnarray}
} where it is possible to distinguish the contributions of the
gaussian integration to the kinetic and potential energy (respectively
the terms proportional to $1/\sigma^2$ and $\sigma^2$).  The
requirement of stationarity of this functional yields the equations
for $\tilde{\psi}$
\begin{eqnarray}
\label{eq:dr-GPE}
i\hbar\partial_t\tilde{\psi}(z)&=&\left\{-\frac{\hbar^2}{2m}
\frac{1}{\lambda_z^2}\nabla^2_z+\frac{\hbar^2}{2m}
\frac{1}{\sigma^2\lambda_\perp^2}+U_{1D}(\lambda_z z,t)+
\frac{1}{\lambda_z\lambda_\perp^2}
\left[\frac{m\omega_z^2(0)}{2}z^2+\frac{m\omega_\perp^2(0)}{2}
\sigma^2+\frac{gN}{2\pi\sigma^2}|\tilde{\psi}|^2\right]\right\}
\tilde{\psi}(z)
%\nonumber
\end{eqnarray}
\end{widetext}
and for $\sigma$ \begin{equation}
\label{eq:sigma}
\sigma(z,t)=a_\perp\sqrt[4]{\lambda_z(t)+2aN|\tilde{\psi}(z,t)|^2},
\end{equation}
where $a_\bot$ is defined through the harmonic frequency at $t=0$ by
$a_\bot\equiv\sqrt{\hbar/m\omega_\bot(0)}$.
From Eqs.\ (\ref{eq:gaussian_width}) and (\ref{eq:sigma}), the radial
width $\Sigma$ of the true wave function $\Psi$ is given by
\begin{equation}
\label{eq:Sigma}
\Sigma(r_z,t)=a_\perp\lambda_\perp(t)\sqrt[4]
{\lambda_z(t)\left(1+2aN|\psi(r_z,t)|^2\right)}.
\end{equation}

By combining Eqs.\ (\ref{eq:dr-GPE}) and (\ref{eq:sigma}) we get a 1D
nonlinear Schr\"odinger equation for $\tilde{\psi}$, depending on the
parameters $\lambda_j$ solution of Eq.\ (\ref{eq:lambda}), that we
call \textit{dynamically rescaled} Gross-Pitaevskii equation
(\emph{dr}-GPE). This equation is energy conserving and 
requires the same computational effort of a simple 1D-GPE,
since the numerical solution of Eq.\ (\ref{eq:lambda}) is
straightforward. 
In the particular case of a time-independent
harmonic potential Eqs.\
(\ref{eq:dr-GPE}) and (\ref{eq:sigma}) reduce to the
NPSE of \cite{npse}. 

The contribution due to the transverse part of the
energy, that has been neglected within the approximation $\nabla^2
\tilde{\phi} \approx\nabla^2_\perp\tilde{\phi}$, can be estimated
by considering the integral
\begin{equation}
\label{ekin_trasc}
\frac{\hbar^2}{2m}\langle|\nabla_{r_z}\phi|^2\rangle=\frac{\hbar^2}{2m}\frac{\langle|\nabla_z\tilde\phi|^2\rangle}{\lambda_z^2}=
\frac{\hbar^2}{2m\lambda_z^2}\int dz\
\frac{\sigma'^2}{\sigma^2}|\tilde{\psi}|^2
\end{equation}
where $\sigma'\equiv\partial_z\sigma$.  We have verified that, in the
case of a general  time-dependent harmonic trapping,
the contribution of this term is a
negligible fraction of the total energy (less than $0.1\%$) in a wide
range of trap anisotropies $\omega_z/\omega_\perp$ and number of atoms
$N$. As we show in the next Section, even in the presence of an optical
lattice along the axial direction, where the density can show deep
modulations on a short scale, the model gives quite accurate results.

In the rest of the paper we consider some
applications of the model and make comparisons with analytical results in the
Thomas-Fermi (TF) limit, full 3D numerical solutions of the GPE and
experimental results, where available.  
The parameters are chosen in the range of typical
experiments at LENS \cite{burger,cataliotti,pedri}. In
particular, we model condensates formed by $2\cdot10^4\leq N \leq
2\cdot10^5$ atoms of $^{87}$Rb in a cigar-shaped configuration with
axial and radial harmonic trapping frequencies respectively
$\omega_z=2\pi\cdot9$ Hz and $\omega_\bot=2\pi\cdot92$ Hz.

To numerically solve the model, we use a Runge-Kutta algorithm for the
scaling equations and a FFT split-step method for the
\emph{dr}-GPE evolution, mapping the wave
function on a discretized lattice \cite{NR2nd,splitstep,sites}. 
The ground state of the system is found by using 
a standard imaginary time evolution
\cite{BEC_review}.

%=====================================================
\section{Applications}
%=====================================================
\subsection{Harmonic trapping}
\label{sec:harmonic}

\begin{figure}
\centerline{\includegraphics[width=8.6cm,clip=,angle=0]{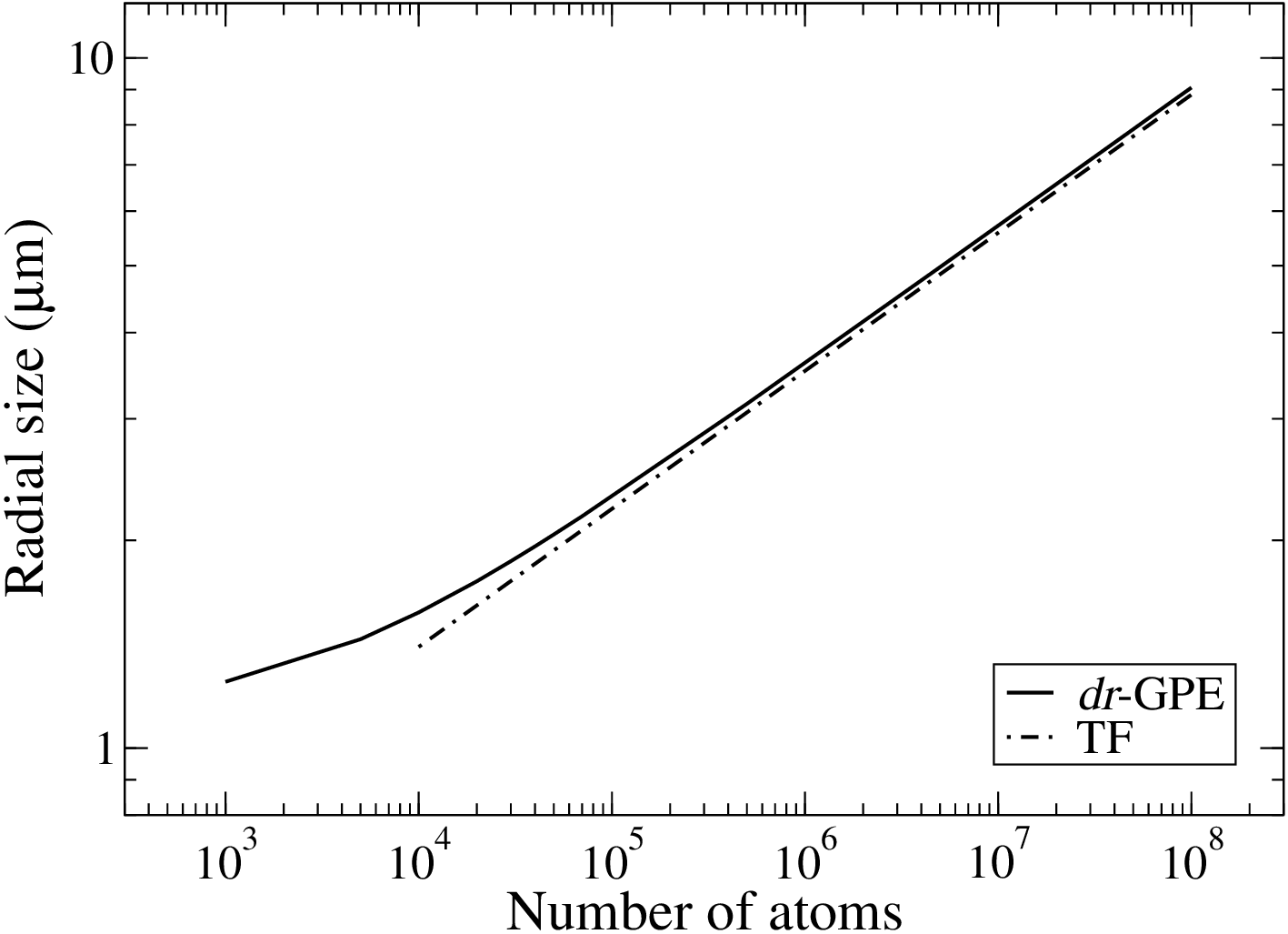}}
\caption{\label{sqrt_r2_senza_reticolo}Comparison of the \emph{dr}-GPE
and TF predictions for the radial size of a condensate in a pure
harmonic potential as a function of the number of atoms N.  The shift
between the two curves is expected since the TF approximation
truncates the tails of the wave function, leading to a systematic
underestimate of the ground state size, especially for small $N$.}
\end{figure}

{We start by considering the case of a condensate confined 
in a pure harmonic potential.} As {discussed} in \cite{npse}, 
the ground state axial
{profile given by the gaussian factorization (\ref{eq:factor}) }
reproduces the full 3D  solution much better
than other known 1D model in both the weakly and strongly (TF)
interacting regimes.  The model also provides
 reasonable results for the radial profile: 
the gaussian Ansatz, though  
slightly overestimating the central density, gives a 
prediction for the rms radius 
$R_{rms}\equiv\sqrt{\langle r_\bot^2\rangle}$ 
really close to the analytical TF result 
even in case of very large number of atoms, 
 as shown in Fig.\ \ref{sqrt_r2_senza_reticolo}.

\begin{figure}
\centerline{\includegraphics[width=8.6cm,clip=,angle=0]{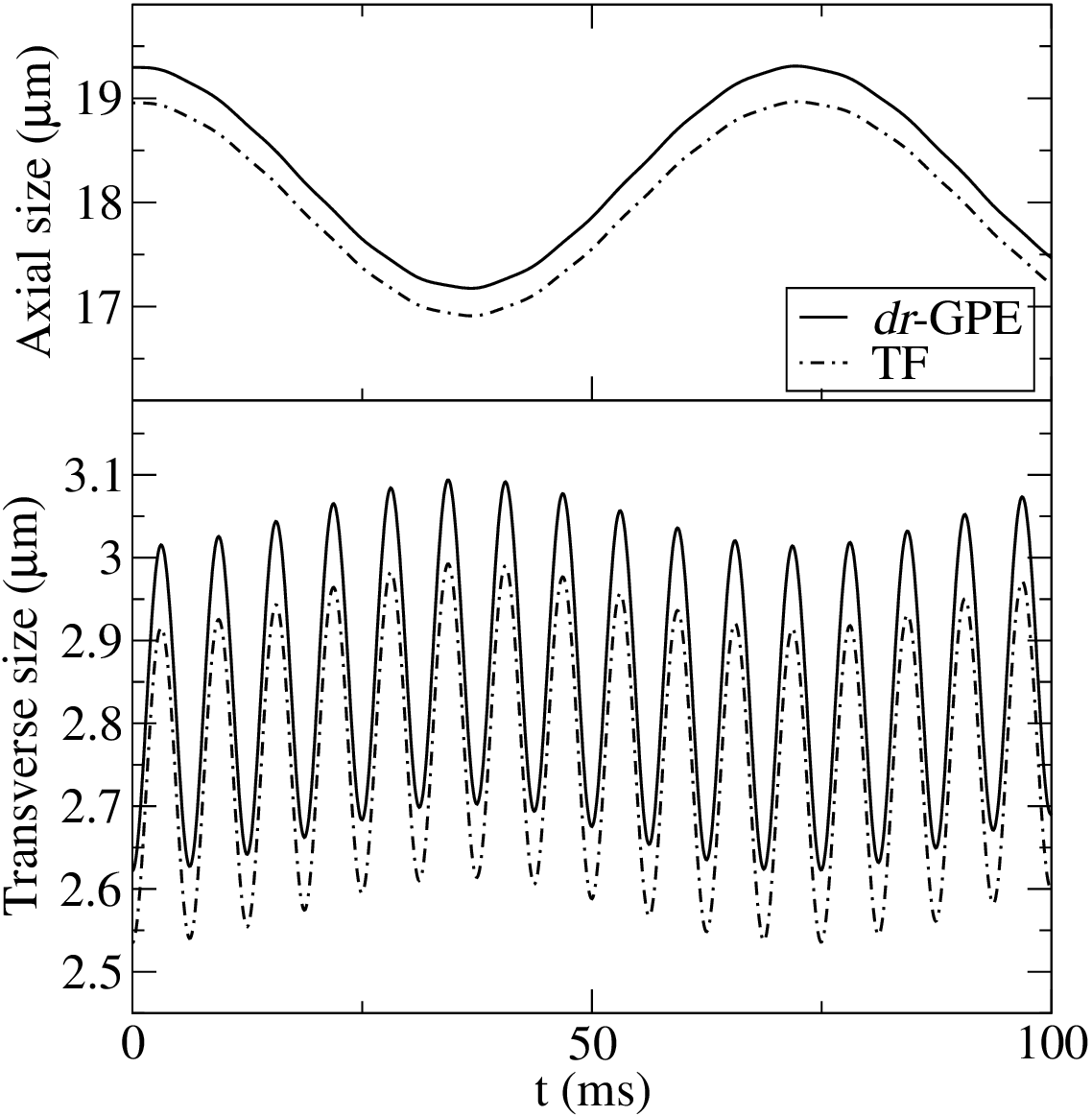}}
\caption{Evolution of the axial ($Z_{rms}$, top) 
and radial ($R_{rms}$, bottom) sizes of a
condensate performing shape oscillations, respectively obtained
solving the \emph{dr}-GPE and the TF Eqs.\ (\ref{eq:lambda}) and 
(\ref{eq:TF-sizes}): in both graphs
it is possible to observe the superposition of the quadrupole and the
faster transverse breathing frequencies ($N=2\cdot10^5$,
$\omega_z=2\pi\cdot8.7$ Hz).}
\label{fig:oscill}
\end{figure}

A remarkable feature of the \emph{dr}-GPE  is the possibility 
to describe the interplay between
longitudinal and transverse oscillations
induced by modulations of both the axial and radial part of the
trapping potential.
To illustrate this aspect we
 consider the low-lying collective excitations induced by
a sudden variation of the transverse external confinement.  In particular
we start with a condensate in the ground state of an anisotropic trap
with $\omega_\bot=2\pi\cdot92$ Hz and, at $t=0$, we suddenly switch
the transverse frequency to a final value of $\omega_\bot=2\pi\cdot80$
Hz, thus inducing shape oscillations involving both the radial and
axial directions.  For very elongated condensates in the TF regime,
these oscillations are characterized by the quadrupole and transverse
breathing frequencies, $\omega^Q=\sqrt{5/2}\omega_z$ and
$\omega^{TB}=2\omega_\bot$ respectively \cite{BEC_review}.  In Fig.\
\ref{fig:oscill} we show the behavior of the radial and axial sizes
(respectively $R_{rms}$ and $Z_{rms}\equiv\sqrt{\langle z^2\rangle}$ ), 
comparing the  \emph{dr}-GPE predictions 
with the TF values.  The agreement is remarkably good
for both the frequency and amplitude of the oscillations; the slight
shift between the two curves is expected, due to the truncation
of the tails of the wave function in the TF approximation
(see Fig.\ \ref{sqrt_r2_senza_reticolo}).  Despite the
1D (axial) character of the \emph{dr}-GPE, Fig.\
\ref{fig:oscill} shows that the
equation well mimics the predicted coupling between the axial and
radial degrees of freedom.  We have checked that the
same behavior can be reproduced by changing the axial
trapping instead of the radial one, or by using a resonant drive.

\begin{figure}
\centerline{\includegraphics[width=8.6cm,clip=,angle=0]{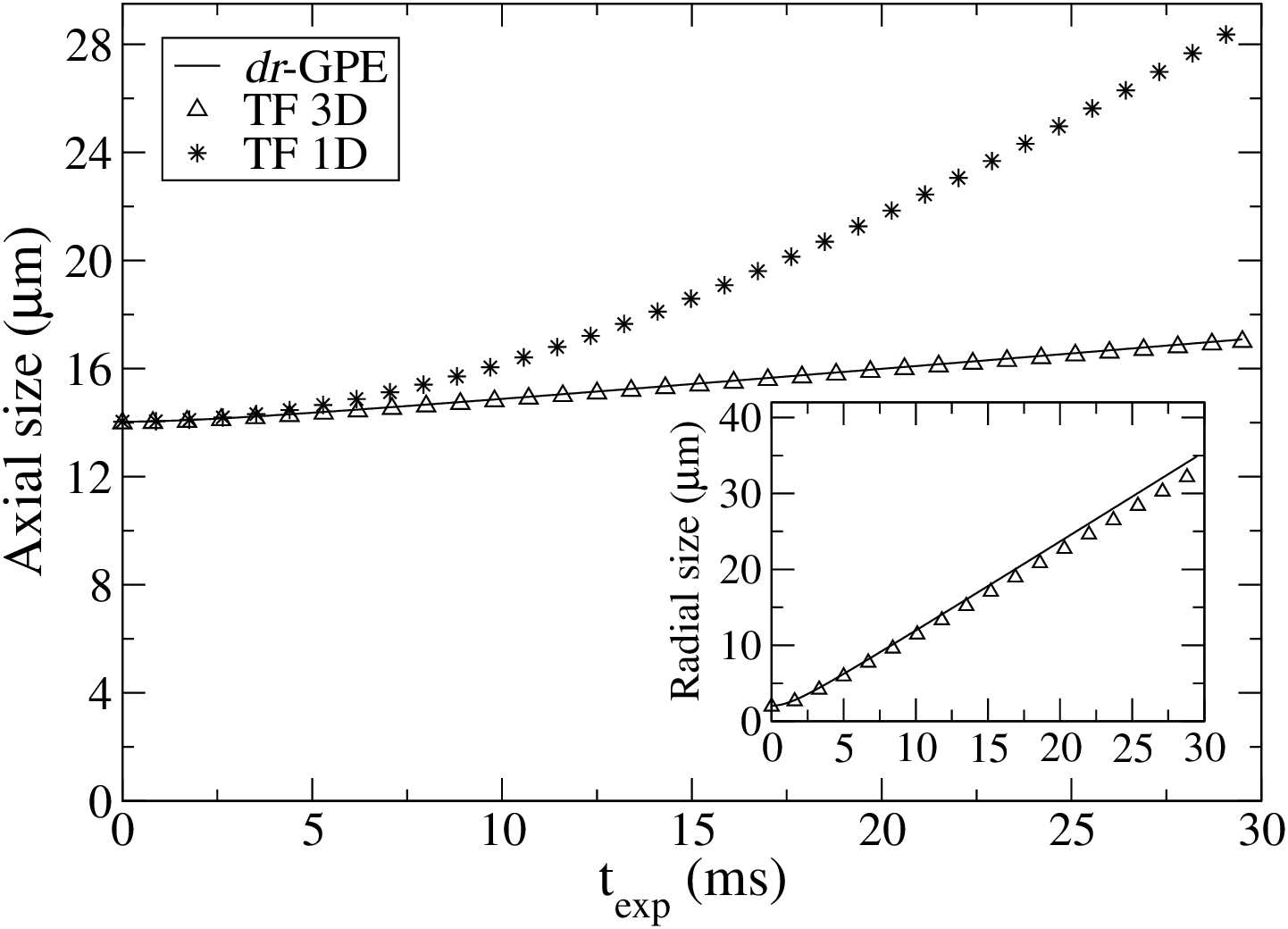}}
\caption{\label{confronto_exp_1D_3D_inset} Axial and radial sizes 
(respectively $Z_{rms}$ and $R_{rms}$) of
the condensate during a free expansion from a pure harmonic potential,
as obtained with the \emph{dr}-GPE and within the TF approximation in
1D and 3D.}
\end{figure}

Another main difference between our model and any static 1D
renormalization of the GPE  (1D-GPE) comes out when we consider the ballistic
expansion of a condensate suddenly released from the trapping
potential.
In fact, in the typical experimental regimes the free expansion 
of a condensate is governed by the
TF equations \cite{BEC_review}
\begin{equation}
\label{eq:TF-sizes}
\sqrt{\langle r_i^2\rangle_{_{TF}}}(t)=\lambda_i(t)\sqrt{\langle
r_i^2\rangle_{_{TF}}}(0)
\end{equation}
where $\lambda_i(t)$ are solutions of Eq.\ (\ref{eq:lambda}) with
$\omega_z(t)=\omega_\bot(t)=0$.  In the case of a
1D GPE a similar relation holds for the axial size, except that now
the only scaling parameter obeys
\begin{equation}
\label{eq:lambda1D}
\ddot{\lambda_{1D}}(t)= \frac{\omega^2_z(0)}{\lambda_{1D}^2(t)}
-\omega_z^2(t)\lambda_{1D}(t).
\end{equation}
One immediate consequence of Eq. (\ref{eq:lambda1D})
is that in the TF limit a 1D GPE equation
obtained through a simple renormalization of the coupling constant,
although suited to reproduce the (axial) ground state of the system,
in general overestimates the repulsive character of the interaction 
during the expansion.
 This is shown in Fig.\ \ref{confronto_exp_1D_3D_inset} where we plot
the axial size of the expanding condensate as obtained from the
\emph{dr}-GPE and the 1D and 3D TF scaling.  
The nice agreement between the \emph{dr}-GPE solution and
the full dimensional case is due to the use of the scaling equations
and to the dynamically reduced nonlinearity 
of the equation (\ref{eq:dr-GPE}) that, in the strongly interacting (TF) limit
$aN|\tilde\psi|^2\gg\lambda_z$, takes the form
{\setlength\arraycolsep{0pt}
\begin{eqnarray}
\label{eq:TF-dr-GPE}
i\hbar\partial_t\tilde{\psi}(z)=&&\left\{-\frac{\hbar^2}{2m}
\frac{1}{\lambda_z^2}\nabla^2_z+U_{1D}(\lambda_z z,t)\right.\\
&&\!\!\!\!\!\!\!\!\!\!\!\!\left.+\frac{1}{\lambda_z\lambda_\perp^2}
\left[\frac{m\omega_z^2(0)}{2}z^2+\frac3 2
\hbar\omega_\bot(0)\sqrt{2aN}|\tilde{\psi}|\right]\right\}
\tilde{\psi}(z)\;.\nonumber
\end{eqnarray}}
The inset of Fig.\ \ref{confronto_exp_1D_3D_inset} shows
that even for the transverse dynamics, despite the 1D character of the
\emph{dr}-GPE, the agreement with the 3D (TF) solution is remarkably
good.

%=====================================================
\subsection{Optical lattice}
\label{sec:optical}
%=====================================================

We now turn to consider the case of a condensate in a 1D optical lattice.
 Experimentally this is realized by means of a
retro-reflected laser beam which creates a periodic
potential of the form
\begin{equation}
U_{1D}(r_z)=s\cdot E_r\cos^2(2\pi r_z/\lambda_{opt})
\end{equation}
where $\lambda_{opt}$ is the wavelength of the laser, 
$E_r\equiv h^2/2m\lambda_{opt}^2$ is the recoil energy of an
atom absorbing one lattice phonon and $s$ is a dimensionless parameter
controlling the intensity of the lattice.
Here we chose $\lambda_{opt}=795$ nm and intensities in the
range $0\leq s\leq6$ as reported in the experiments 
\cite{burger,cataliotti, pedri}.

\begin{figure}
\centerline{\includegraphics[width=8.6cm,clip=,angle=0]{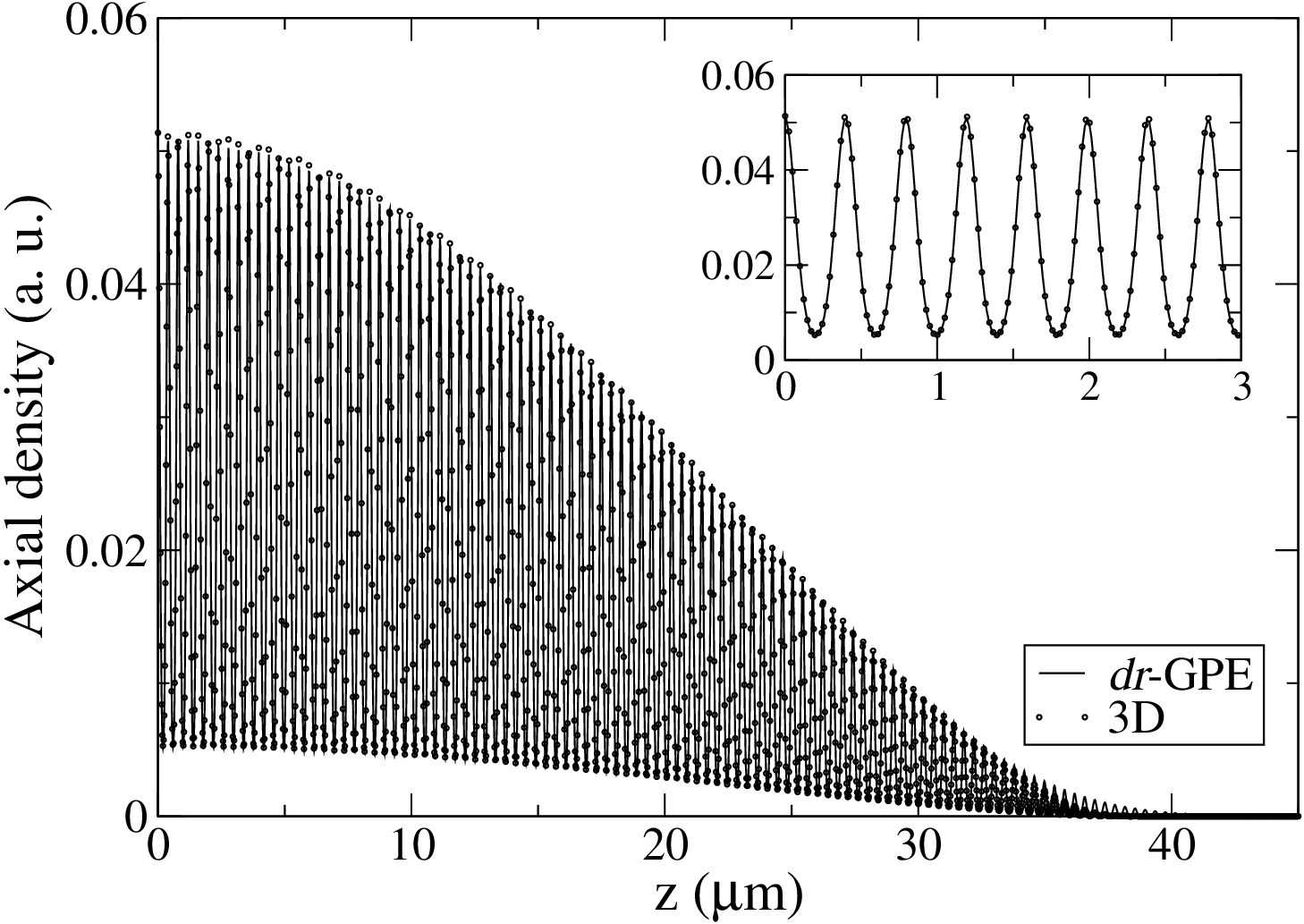}}
\centerline{\includegraphics[width=8.6cm,clip=,angle=0]{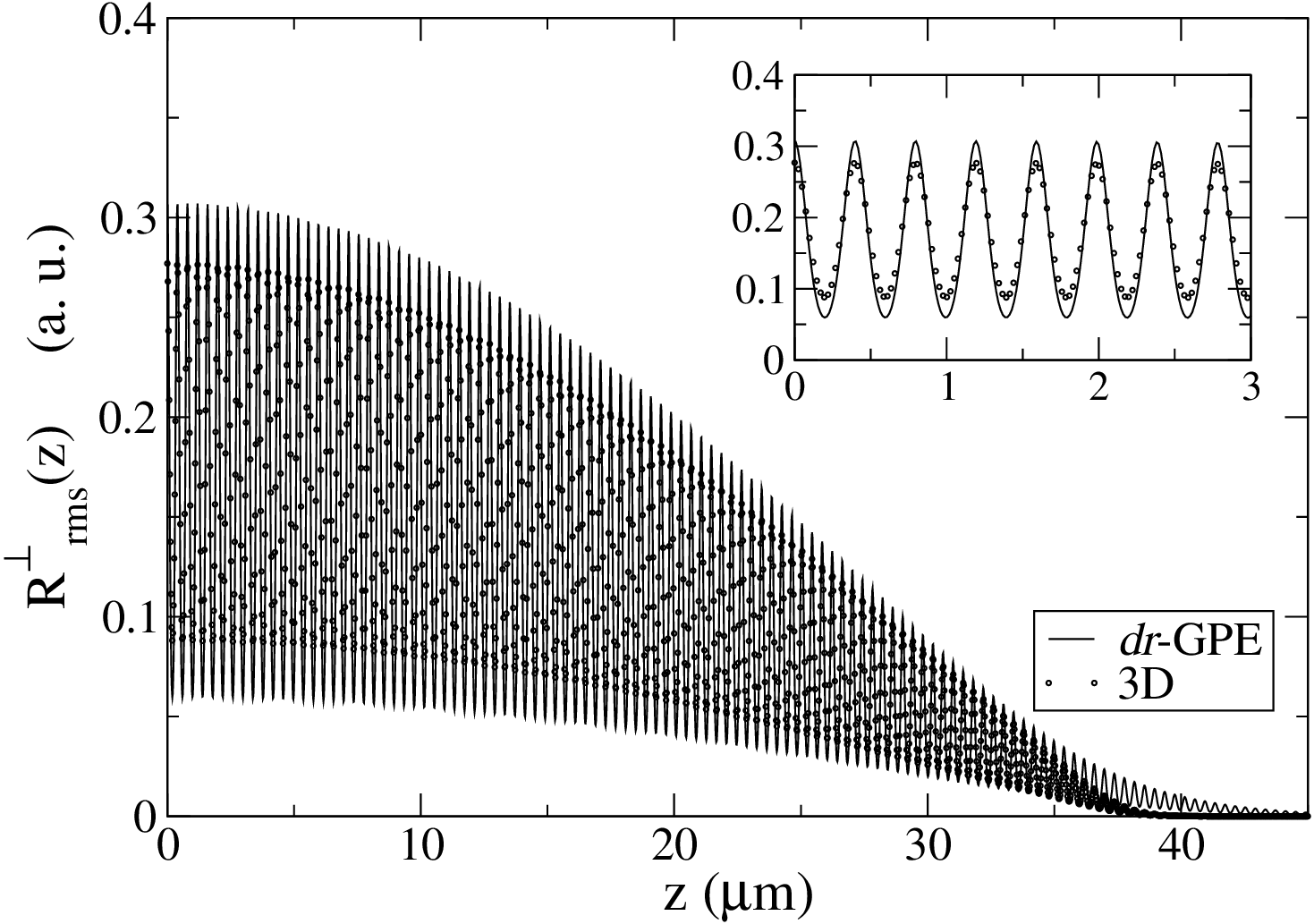}}
\caption{\label{profilo_assiale}Axial density $\rho(z)$ (top) and rms
transverse radius, averaged only on the transverse coordinates and
plotted as function of the axial coordinate (bottom); the insets show
the central region of the condensate ($N=5\cdot10^4$, $s=5$).}

\end{figure}

The ground state of the system is characterized by 
deep periodic variations in the axial density and by pronounced
modulations in the radial width,
that are well reproduced
by our model even for high values of lattice
intensities.  This is evident in Fig.\ \ref{profilo_assiale} 
 where we compare the axial density
\begin{equation}
\rho(z)\equiv\langle|\Psi|^2\rangle_\bot=\int\!d^2r_\bot\ |\Psi|^2
\end{equation}  
and  the rms transverse radius, integrated only over the radial directions
\begin{equation}
R^\perp_{rms}(z)\equiv
\sqrt{\langle r^2_\bot\rangle_\bot}=\sqrt{\int\!d^2r_\bot\
r^2_\bot|\Psi|^2}\ ;
\end{equation}
with the full 3D solution.

To address the issue of collective excitations we consider
both dipole oscillations induced by a sudden displacement of the 
harmonic potential (here we use $\Delta z=5\ \mu$m),
and quadrupole modes excited as before by changing the harmonic
transverse confinement. 

Concerning the former, on general grounds one expects a frequency shift
which can be explained in terms of a mass renormalization due to the
{\em s}-dependence of the dispersion relation in the lowest band of the 
periodic potential \cite{ashkroft,burger}:
$\omega^D_z\rightarrow\sqrt{m/m^*}\omega^D_z$.  In a recent paper it
has been predicted, by using a tight binding Ansatz, that the same
effective mass $m^*$ can account also for the modification of the
quadrupole frequencies \cite{kraemer}: for elongated
condensates in the TF regime one expects a linear relationship between the two
frequencies, $\omega^Q_z=\sqrt{5/2}\omega^D_z$, independently of the
lattice intensity $s$.

\begin{figure}
\centerline{\includegraphics[width=8.6cm,clip=,angle=0]{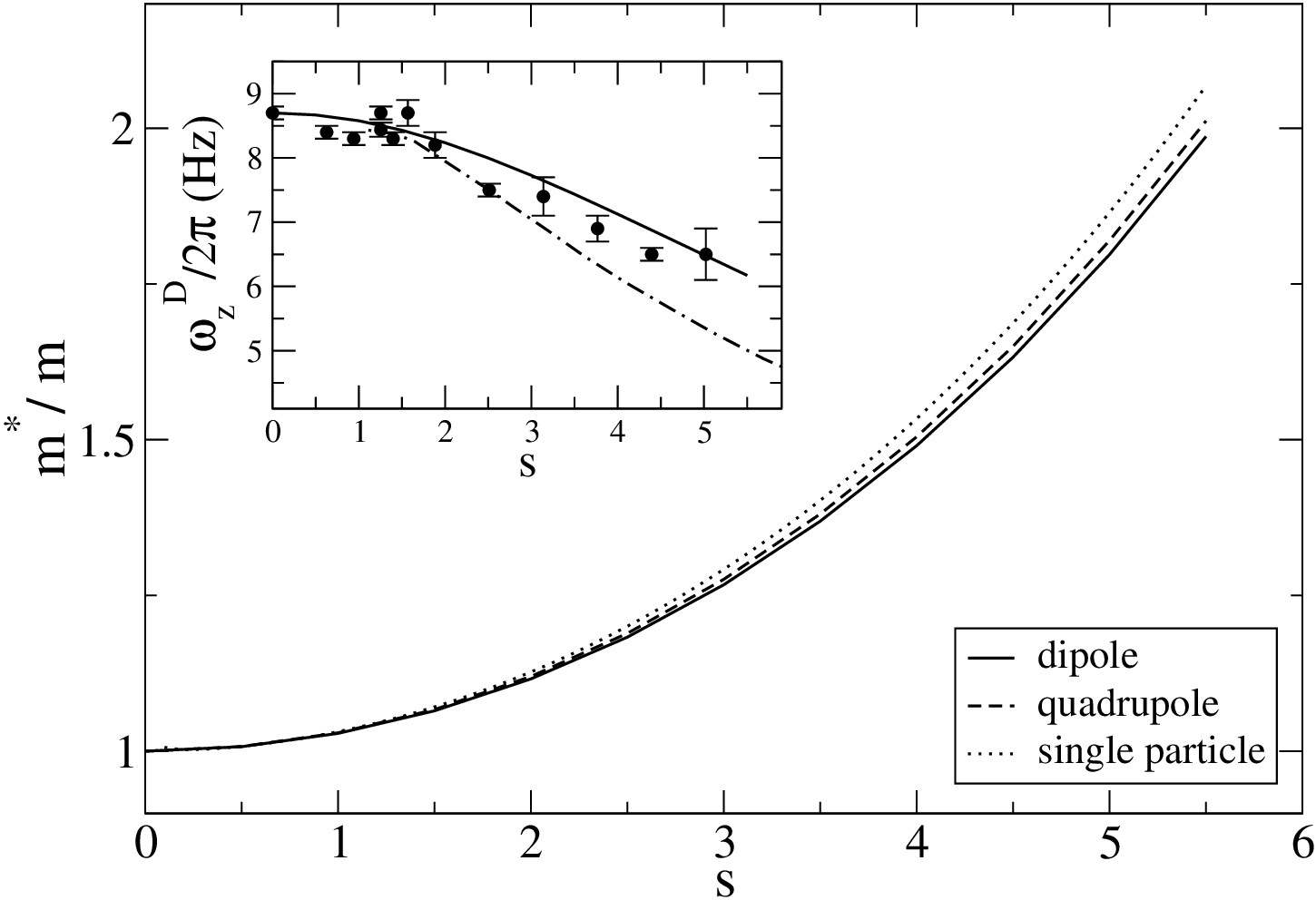}}
\caption{Effective mass $m^*$, normalized to the unperturbed value
$m$, as a function of the lattice intensity $s$, obtained respectively
from the dipole (continuous line) and quadrupole (dashed) frequencies,
and compared to the case of a single particle in a periodic 
potential (dotted). In the inset the frequency of the dipole mode 
is compared with the  experimental data and 
theoretical curve (dashed-dotted) from \cite{cataliotti}.}
\label{fig:mass}
\end{figure}

In Fig.\ \ref{fig:mass} we compare the effective mass extracted from
the dipole and quadrupole modes with that of  a single
particle in a periodic potential, near the Brillouin zone center
\cite{ashkroft}. This picture shows that the predicted relation between
$\omega^Q_z$ and $\omega^D_z$ is very well verified also in the
weak binding regime of small $s$, and that the effects of mean
field interaction and of harmonic trapping (both included in the 
\emph{dr}-GPE description) give only a minor correction to
the single particle picture in a uniform periodic potential.
Our results are also in good agreement with a recent experiment
 realized at LENS \cite{cataliotti}, as shown in the inset of 
Fig.\ \ref{fig:mass}.

\begin{figure}
\centerline{\includegraphics[width=8.6cm,clip=,angle=0]{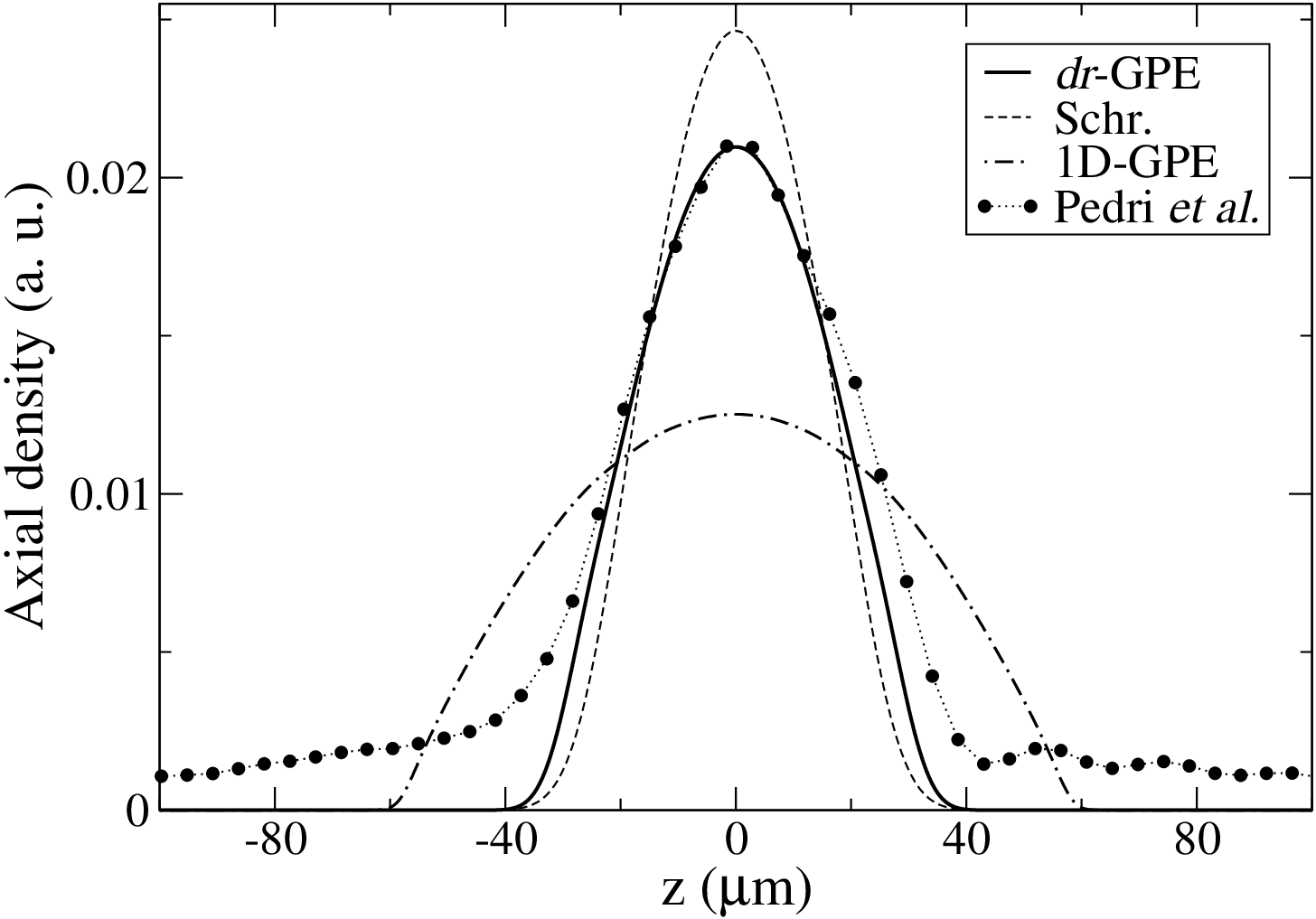}}
\caption{\label{confronto_centr}Axial density of the central peak, as
obtained respectively with the \emph{dr}-GPE, a Schr\"odinger
propagation and the 1D-GPE introduced in \cite{tripp}, compared with
the experimental distribution of \cite{pedri} 
(the wings are due to the presence of a small thermal component), after
$t_{exp}=29.5$~ms.}
\end{figure}

Other interesting features of BECs confined in 1D optical lattices
are the interference effects which take place after the removal of
the combined optical+harmonic trapping potential.
In fact, the free expansion of the system causes the single condensates
forming the initial array to overlap and leads to the formation of
neat interference patterns, showing up as lateral peaks moving with
quantized speeds $v_n=\pm n\cdot2h/\lambda_{opt} m$ (with $n$
integer) \cite{pedri}.  
This is essentially a 1D effect, which reflects the momentum
distribution of the ground state along the axial direction,
and it is well reproduced by our model, regarding both the position
and the population of the first lateral peaks.
An interesting point concerns the axial expansion of the central peak.
A careful analysis of the expanded density profile shows that,
compared to other existing 1D equations, the \emph{dr}-GPE gives a
better estimate of the axial size of the central peak.
  In fact, as can be seen in Fig.\ \ref{confronto_centr}, a (free)
Schr\"odinger propagation of the ground state underestimates the axial
width, while the statically renormalized 1D-GPE introduced in
\cite{tripp}, though appropriate to describe the ground state, largely
overestimates the axial size after the expansion, as discussed in the
previous section. This is indeed due to the fact that the
meanfield term of the \emph{dr}-GPE has a linear character 
($\propto|\tilde\psi|$, see Eq. \ref{eq:TF-dr-GPE}) 
which lies between the Schr\"odinger 
($\propto|\psi|^0$) and the GPE ($\propto|\psi|^2$) cases.
Our model gives accurate predictions also for the radial expansion 
of the central peak, which is essentially governed by the TF scaling
solution, in agreement with the experimental values and 
the theory of \cite{pedri}.

We conclude this section by discussing a time-of-flight method that
could be used to probe the dispersion relation in the presence of
a periodic potential \cite{probe}, 
allowing for the investigation of negative-mass effects
during the expansion in an optical lattice.
The idea is the following: we start with a condensate
in the ground state of a pure harmonic potential, and we accelerate it 
up to a given velocity $v=\hbar q/m$ in the axial direction,
{\em e.g.} by exciting a dipole motion in the harmonic potential.
Then we switch off the trap and adiabatically ramp an optical 
lattice along the longitudinal direction up to a desired height $s$.
Here we consider the case of a linear ramp of $15$ ms,
which is sufficient to ensure the adiabaticity of the process \cite{band}
for the velocities considered here.
We have indeed verified that applying the reverse ramp after the first one,
the additional Fourier components of the wave function at $q\pm2lq_B$
(with $l$ integer) disappear.

\begin{figure}
\centerline{\includegraphics[width=8.6cm,clip=,angle=0]{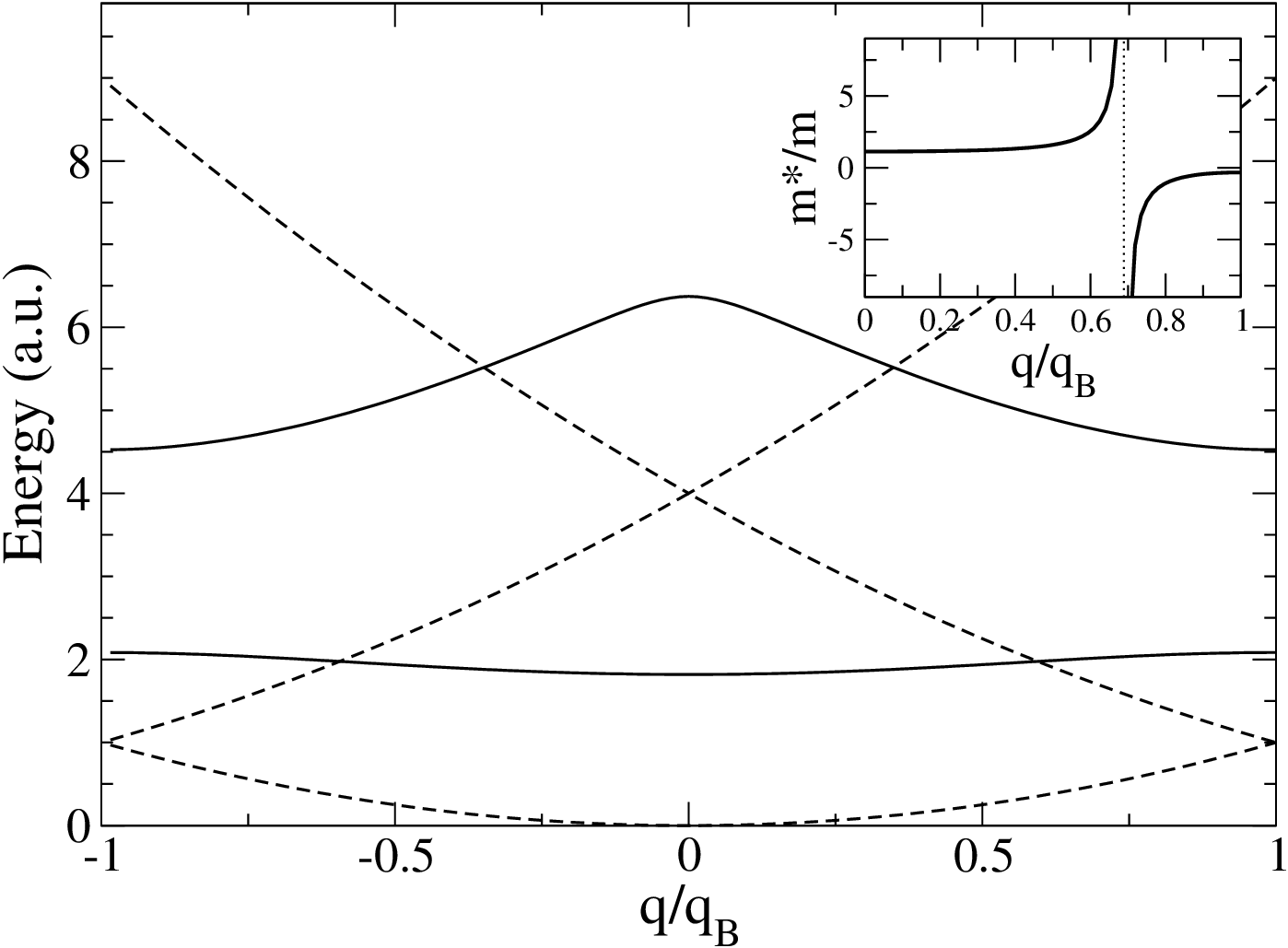}}
\caption{Structure of the lowest bands for a single particle in a 1D
periodic potential with $s=2$ (solid lines), compared to the
free particle case (dashed). The inset shows the behavior of the 
effective mass $m^*$, inversely proportional to the band curvature, 
in the first band for $s=2$.
\label{fig:banda}}
\end{figure}

With this procedure we can load the condensate in a Bloch state of
axial quasimomentum $\hbar q$ and band index $n$, 
and therefore explore effects of
effective mass (which depends on $s$, $n$ and $q$) far from the Brillouin 
zone center, during the expansion of the system.
In particular, since meanfield effects are ruled out after few milliseconds, 
the radial and axial expansion are almost decoupled, and the latter can 
be accounted for by the single particle mass renormalization, 
as shown in Fig. \ref{fig:banda} \cite{rolston,ashkroft}. 
The advantage of loading the condensate in the lattice only after the 
acceleration procedure is that
in the reverse case the occurrence of dynamical instability effects
prevents the possibility of accelerating the condensate above
a critical velocity $v_c<v_B$  \cite{mod_inst}, where
$v_B=\hbar q_B/m=h/m\lambda$ is the Bragg velocity. 

\begin{figure}
\centerline{\includegraphics[width=8.6cm,clip=,angle=0]{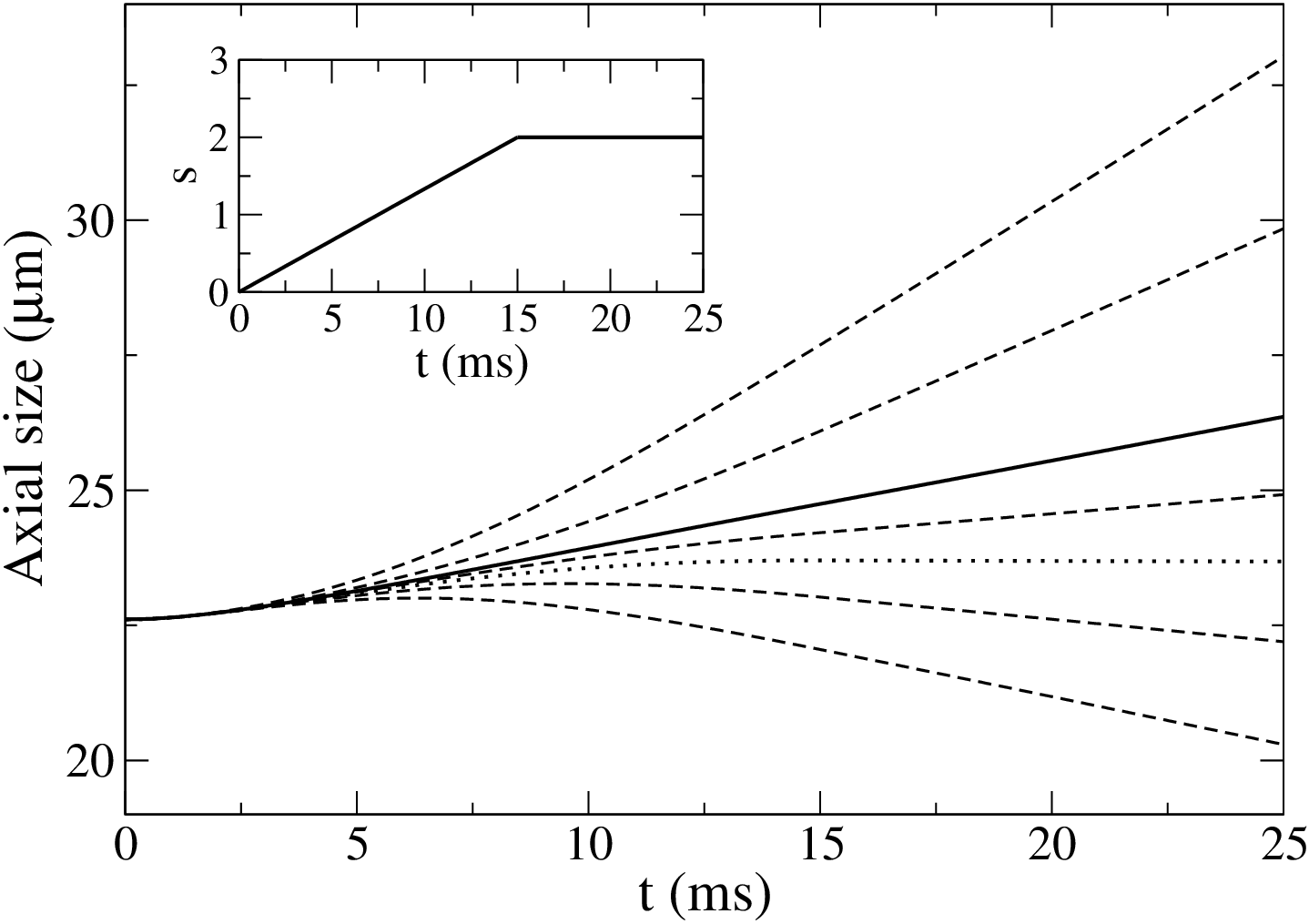}}
\caption{Evolution of the axial width of a condensate which is loaded
during the expansion in a 1D optical lattice, for
various velocities: $q/q_B=1.3,1.2,0$ (continuous line), $0.4,0.685$
(dotted), $0.75,0.8$, respectively from top to bottom. 
The lattice intensity is ramped linearly 
in the first $15$~ms of expansion, and then kept fixed at the value $s=2$
for other $10$~ms (see the inset).
\label{fig:axial}}
\end{figure}

In Fig. \ref{fig:axial} we show the evolution of the axial
size of the condensate during the expansion in the optical lattice,
for various initial velocities $v$. The behavior of the system is 
essentially determined by the single particle effective mass
which characterizes the diffusive (kinetic) term in the evolution
equation (as stated before, meanfield effects become negligible after 
few milliseconds of expansion). Indeed, the figure shows that when
$q$ approaches the value $\tilde{q}=0.6895~q_B$ at which the single
particle effective mass for $s=2$ becomes infinite (see Fig. \ref{fig:banda}), 
the expansion in the direction
of the lattice gets frozen. For values between   $\tilde{q}$ and
$q_{B}$, $m^*$ becomes negative, and the condensate starts 
contracting. On the contrary, when the condensate is loaded
in the second band, that is for $q>q_{B}$, the expansion is enhanced
due to the strong positive curvature near the second band edge.

\begin{figure}
\centerline{\includegraphics[width=8.6cm,clip=,angle=0]{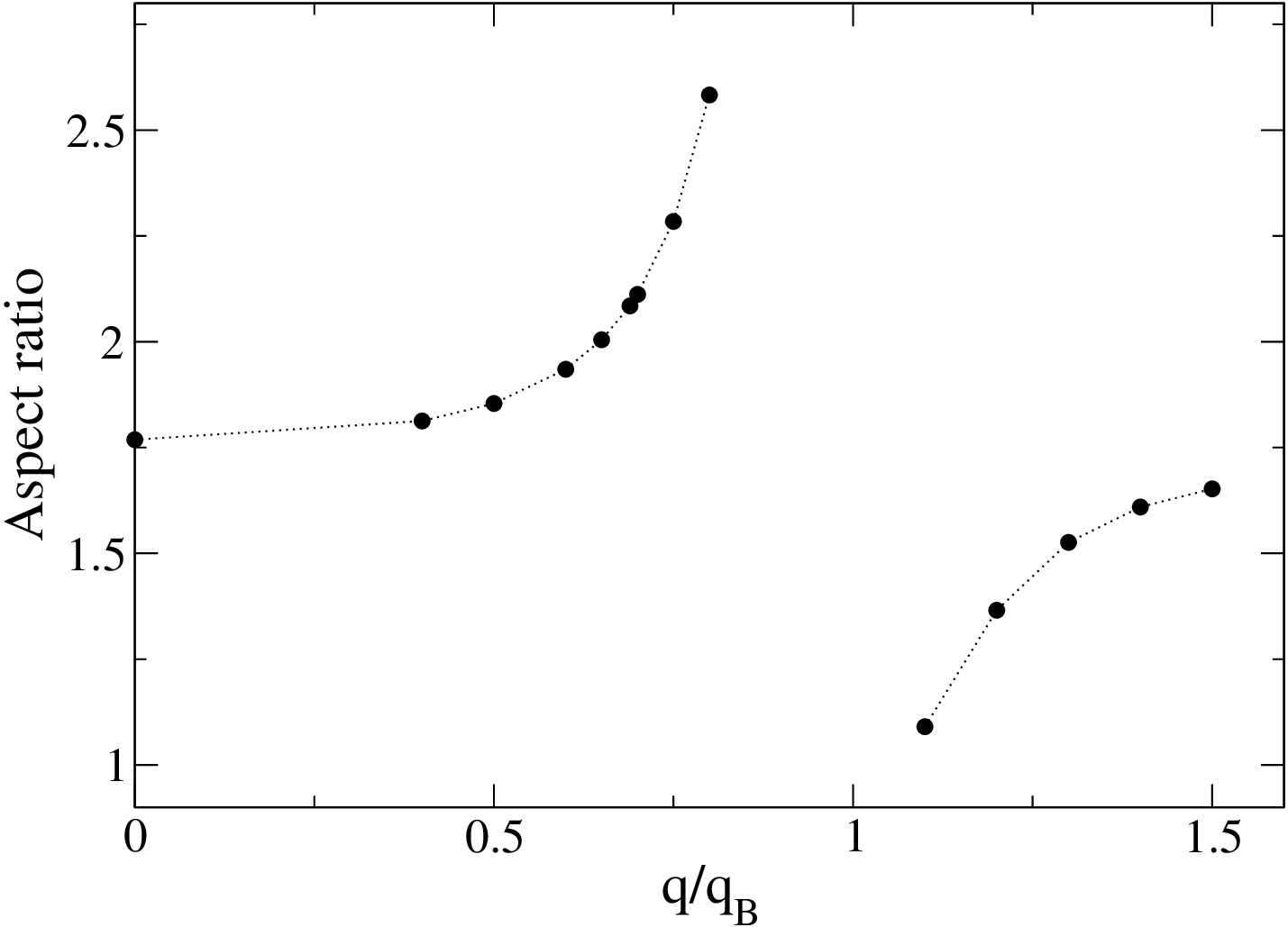}}
\caption{Aspect ratio after $25$~ms of expansion, for a condensate loaded
adiabatically in a 1D optical lattice of height $s=2$, as a function
of the condensate velocity.
\label{fig:aspect}}
\end{figure}

Since the transverse expansion is only weakly affected by the presence of the 
lattice, these dramatic effects on the axial expansion can be directly
measured on the aspect ratio after a fixed
evolution time. Indeed, as shown in Fig. \ref{fig:aspect}, 
the aspect ratio is characterized by a finite jump across
the first Brillouin zone boundary, which reflects the discontinuity
in the band curvature (the effective mass passes from a negative to
a positive value).
With the same technique one could probe the band structure also for
higher values of $q$ and/or $s$. In the latter case a longer ramp time 
could be required in order to ensure an adiabatic loading \cite{band}.

%=====================================================
\section{Conclusions}
\label{sec:conclusions}
%=====================================================

We have presented an effective 1D model for the evolution of an
elongated Bose-Einstein condensate in time dependent potentials
whose radial component is harmonic, in the presence of an arbitrary
axial potential. The model exploits the scaling and gauge
transformations of \cite{castindum,kagan}, that reabsorb
most of the evolution due to the time-dependent variation of the
harmonic  potential. This makes possible a gaussian
factorization for the radial part of the condensate wave function,
even in case of a free expansion after the release from the trap.

The evolution of the system is described by a nonlinear 1D wave
equation ({\em dr}-GPE) for the axial wave function,
dynamically rescaled by means of the aforementioned transformations.
Despite the 1D character of the {\em dr}-GPE, the model 
also accounts for the radial dynamics of the system, that is necessary
to correctly describe collective oscillations and the free expansion
of the condensate.

The accuracy of the model has been tested by comparing its predictions
with the TF solution in case of pure harmonic trapping and
with recent experimental and theoretical results for the dynamics and
the expansion of BECs in 1D optical lattices.
We have also shown that the quasimomentum dependence of the effective mass 
has dramatic consequences, which could be easily accessed in the experiments
by accelerating and letting expand a condensate in an optical lattice.

\begin{acknowledgments}
We are grateful to the group at LENS for providing us with the
experimental data and for stimulating discussions. We thank C. Fort
for a careful reading of the manuscript. P. M. acknowledges fruitful
discussions with N. Piovella and has been supported by the EU under
Contract No.\ HPRI-CT 1999-00111.
\end{acknowledgments}


\begin{thebibliography}{}
\bibitem{sites} 
The typical number of lattice sites
used is 512 for the dynamics and the expansion in a pure harmonic
potential, 4096 for the trapped dynamics in the presence of an optical
lattice and 32768 for the expansion of the coherent array of
condensates.
\end{thebibliography}
\end{document}